\documentclass[12pt]{iopart}
\expandafter\let\csname equation*\endcsname\relax
\expandafter\let\csname endequation*\endcsname\relax
\usepackage{microtype}
\usepackage{float}
\usepackage{graphicx}
\usepackage{amsmath}
\usepackage{booktabs}
\usepackage{amssymb}
\usepackage{amsfonts}
\usepackage{color}
\usepackage{hyperref}

\begin{document}

\title{Entropy-Deformed Hamiltonian Dynamics of Schwarzschild Black Holes: A Superstatistical Approach}

\author{O.Garcia, O.Obregón, J. Ríos Padilla}

\address{University of Guanajuato, División de Ciencias e ingenierías, campus León, CP:37150, Méx.}
\ead{jdj.riospadila@gmail.com}
\vspace{10pt}
\begin{indented}
\item[]August 2017
\end{indented}

\begin{abstract}
We study the effective dynamics of the Schwarzschild black hole interior by introducing entropic deformations derived from generalized superstatistical entropies $S_{+}$ and $S_{-}$. The resulting modified Hamiltonians $\bar{H}_{\pm}$, formulated in Ashtekar--Barbero variables, encode quantum gravity-inspired corrections that become significant near the Planck scale. Analytical solutions show that these corrections regularize the classical singularity, replacing it with a finite anisotropic core characterized by bounded canonical variables and a minimal internal area. For $S_{-}$ ($\alpha_{-} > 0$), curvature invariants remain finite, yielding a completely regular interior, whereas $S_{+}$ ($\alpha_{+} < 0$) leads to a localized region of high curvature associated with a cigar-like throat. The interior and exterior geometries are thus connected through this high-curvature region, indicating that the classical singularity is replaced by an entropic transition layer. These features reproduce loop quantum gravity phenomenology without invoking polymer discretization.
 \end{abstract}

%
%
%
%
%

\section{Introduction}

As it is well known, Black Holes emerge as a particular solution of Eintein´s field equations. A quantum gravitational framework is required to understand the behavior of Black Holes at Planckian scales. This has renewed interest in exploring the quantum aspects of black holes to gain insights into the underlying structure of spacetime.

Among non-perturbative approaches to quantum gravity, Loop Quantum Gravity (LQG) introduces the Ashtekar-Barbero variables, where the connection $A^i_a$ and the densitized triad $\xi^a_i$ define a classical phase space suited for quantization\cite{Thiemann_2007}. In the context of spherically symmetric spacetimes, such a phase space describes the black hole interior as a Kantowski-Sachs cosmology\cite{Kelly_2020,Gambini_2020,Ashtekar2006,Bodendorfer_2021,bojowald2020blackholemodelsloopquantum}, enabling a Hamiltonian treatment through a suitable choice of lapse function $N(T)$.

Alternative quantization techniques inspired by LQG, such as polymer quantization, also provide effective descriptions of quantum geometries\cite{Ashtekar_2018}. Furthermore, the Generalized Uncertainty Principle (GUP) offers a complementary approach by deforming the classical Poisson structure. This deformation leads to the resolution of the singularity through a bounce in the conjugate momentum associated with the 2-sphere radius\cite{Bosso2021}.

An independent but related development arises from non-extensive entropy measures, specifically $S_{+}$ and $S_{-}$, constructed within the framework of superstatistics developed by Beck and Cohen \cite{beck2003superstatistics,PhysRevE.78.051101}. These entropies are derived by considering systems with fluctuating intensive parameters, such as inverse temperature $\beta$, which follow $\Gamma$-distribution. The resulting statistical mechanics naturally lead to modified exponential and logarithmic functions, and ultimately to effective probability distributions \cite{Bizet_2023,MartinezMerino2022}. From these entropic constructions, it is possible to derive effective Hamiltonians $\bar{H}_{\pm}$, expressed as corrections to the classical Hamiltonian $H=\frac{p^{2}}{2}+V(q)$, defined as a low-energy Hamiltonian. These corrections are relevant in the quantum gravity regime and depend on the structure of the entropy measures\cite{tsallis2003constructing,obregon2010superstatistics}. Remarkably, the analysis in \cite{Bizet_2023} shows that the entropies $S_{\pm}$ lead directly to a Generalized Uncertainty Principle (GUP), with deformation parameters of opposite sign: $\alpha_{+}<0$ for $S_{+}$	
 and $\alpha_{-}>0$ for $S_{-}$. This duality has profound implications: while $S_{+}$	corresponds to scenarios with a maximum momentum and no minimal position uncertainty, $S_{-}$ implies a minimal length scale without constraining momentum. Consequently, the effective Hamiltonians $\bar{H}_{\pm}$
  encapsulate these distinct quantum gravity effects, yielding modified commutation relations $[q,p_{\pm}]$ that reproduce GUP from purely entropic principles. When formulated in Ashtekar–Barbero variables, these Hamiltonians provide a natural setting to analyze the corrected dynamics of the black hole interior, where the sign of the deformation parameter encodes repulsive ($S_{+}$) or attractive ($S_{-}$) quantum corrections, potentially linked to singularity resolution.

In this work, we analyze the effective dynamics introduced by the Hamiltonians $\bar{H}_{\pm}$, for the Schwarzschild black hole interior. By solving the associated dynamical equations analytically, we are able to track the full evolution of the canonical variables in Ashtekar–Barbero variables and to reconstruct the corresponding spacetime geometry. Our analysis reveals that the entropy-induced corrections lead to a regularization of the interior: the classical curvature singularity is resolved and replaced by a finite, anisotropic core. In this regime, the areal radius of the 2-sphere collapses to zero, while the radial metric components remain finite, producing a cigar-like structure at the core. This anisotropic behavior reflects the fact that entropy-driven corrections break the isotropic scaling of the Schwarzschild interior, introducing a preferred direction consistent with other loop-inspired treatments of black hole interiors, signaling a genuine semiclassical resolution of curvature singularities. A smooth signature change is also observed across a critical radius $t^{*}$, where the temporal and radial metric components exchange their causal roles, allowing a continuous extension beyond the classical singularity. These features are consistent with loop-inspired approaches to black hole interiors \cite{Ashtekar2006,Bojowald_2005,modesto2006loop}, but here they emerge from a statistical-mechanical framework grounded in generalized entropy rather than polymer discretization. Thus, curvature invariants remain bounded even when the 2-sphere collapses, indicating that the singularity is dynamically resolved. Importantly, the specific entropy measure determines the character of the anisotropy through the sign of the deformation parameter $\alpha_{\pm}$: for $S_{-}$ ($\alpha_{-}$>0), the dynamics incorporates a minimal length scale, leading to an attractively confined anisotropy where the collapse is stabilized in the radial sector and the angular directions dominate the contraction. In this case, although the 2-sphere area $A_{\theta\phi}$ vanishes, the mixed areas 
$A_{r\theta}$ and $A_{r\phi}$ remain finite, signaling that the longitudinal directions preserve a nonzero extension at the core. in contrast, for $S_{+}$ ($\alpha_{+}$	
 <0), the existence of a maximal momentum induces a repulsively regulated anisotropy, protecting the radial direction and producing a cigar-like core: the angular area $A_{\theta\phi}$ again collapses, but the residual areas $A_{r\theta}$ and $A_{r\phi}$, acquire values controlled by $\vert\alpha_{\pm}\vert$, which prevent the complete degeneration of the longitudinal geometry. These two entropy-induced regimes highlight that the resolution of the singularity is accompanied by the emergence of anisotropic cores with distinct geometric imprints, where the persistence or suppression of the areas $A_{r\theta}$, $A_{r\phi}$ and $A_{\theta\phi}$ serves as a direct diagnostic of the underlying entropic deformation.

The article is structured as follows: In the Section \ref{sec:superstatistics}, introduces the derivation of the $S_\pm$ entropies and the respective resulting Hamiltonians. In Section \ref{sec:dynamics},  we present the effective dynamics inside the black hole in terms of Ashtekar-Barbero variables and analyze the geometric nature of these solutions. A brief review of the formalism is considered. In the Section \ref{Conclusions}, we present our conclusions and perspectives from this work.
\section{Superstatistics and effective Hamiltonians}\label{sec:superstatistics}
In this section, we present a concise derivation of the generalized entropies $S_{+}$	
  and $S_{-}$,
 within the superstatistics framework. From these, we obtain modified Hamiltonians $\bar{H}_{\pm}$, incorporating corrections relevant at high energies. 
\subsection{Superstatistics entropies \texorpdfstring{$S_{+}$}{S+} and \texorpdfstring{$S_{-}$}{S-}}
In the framework of the superstatistics\cite{beck2003superstatistics,PhysRevE.78.051101}, a distribution function $f(\beta)$ must be considered as a starting point. In this form, the Boltzmann factor $B(E)$, which contains all the statistical information of the system, can be determined. The Boltzmann factor $B(E)$ is given by:
\begin{equation}
    B(E)=\int_{0}^{\infty}d\beta f(\beta) e^{-\beta E},\label{Eq.1}
\end{equation}
where $E$ is the energy of each microstate with probability $p$. Note that the distribution function $f(\beta)$ must be such that it behaves well in the half space $\beta\in [0,\infty)$, that is, that the integral representation of $B(E)$ converges to the open interval. Some distribution functions $f(\beta)$ that meet the above condition can be found in \cite{PhysRevE.78.051101}. In particular, we focus our attention on distribution functions of the type $\Gamma$ (or $\chi^{2}$), which explicitly depend on the temperature inverse $\beta$ and the probability of $p_{l}$ microstates. The dependence of $f(\beta)$ is given by:
\begin{equation}
    f_{p_{l}}(\beta)=\dfrac{1}{\beta_{0}p_{l}\Gamma(\frac{1}{p_{l}})}\left(\dfrac{\beta}{p_{l}\beta_{0}}\right)^{\frac{1-p_{l}}{p_{l}}}e^{-\frac{\beta}{p_{l}\beta_{0}}},\label{Eq.2}
\end{equation}
where $\beta_{0}$ is the average amount of the inverse of temperature. The Boltzmann factor $B_{p_{l}}(E)$ given by for Eq.(\ref{Eq.1}) associated with the distribution function $f_{p_{l}}(\beta)$ in Eq.(\ref{Eq.2}) is of the form:
\begin{equation}
    B_{p_{l}}(E)=\left(1+p_{l}\beta_{0}E\right)^{\frac{1}{p_{l}}}. \label{Eq.3}
\end{equation}
When the Boltzmann factor $B_{p_{l}}(E)$ is expanded as a power series, which is possible when the dimensionless quantity $\beta_{0}E\ll 1$ we find:
\begin{equation}
    B_{p_{l}}(E)\approx e^{-\beta_{0}E}\left[1+\frac{1}{2}p_{l}\beta_{0}^{2}E^{2}-\dfrac{1}{3}p_{l}^{2}\beta_{0}^{3}E^{3}+\dotsb\right].\label{Eq.4}
\end{equation}
The leading term in the expansion reproduces, in essence, the standard Boltzmann--Gibbs statistics, with the subtle distinction that the temperature $T$ (or inverse temperature $\beta$) must be identified with the averaged temperature $T_{0}$ (respectively $\beta_{0}$). The subleading contributions depend on powers of the dimensionless combination $\beta_{0} E$, which remain small within the regime of validity of the approximation employed. Remarkably, these higher-order terms are universal for a broad class of admissible distribution functions $f(\beta)$, reflecting corrections that are insensitive to the detailed shape of the underlying distribution. The conventional Boltzmann factor is recovered in the particular case where $f(\beta)$ reduces to a Dirac delta distribution, $f(\beta)=\delta(\beta-\beta_{0})$.

Now, to proceed towards the construction of the $S_{+}$ and $S_{-}$ entropy measures, we follow the procedure shown in \cite{tsallis2003constructing,obregon2010superstatistics}. We define the entropy $S=\kappa_{B}\sum_{l=1}^{\Omega}s(p_{l})$, in terms of the generic function $s(p_{l})$, where $p_{l}$ is a parameter for the arbitrary moment. Note, for example, that when $s(x)=-x\ln(x)$, the Boltzmann entropy is recovered. As shown in \cite{tsallis2003constructing}, it is possible to express the generic function $s(x)$ and an internal energy function $u(x)$ in terms of the function $E(y)$ which is obtained from the Boltzmann factor $B(E)$. This means that $s(x)$ and $u(x)$ can be written as:
\begin{equation}
    s(x)=\int_{0}^{x}dy\dfrac{\alpha+E(y)}{1-\dfrac{E(y)}{E^{*}}},\qquad u(x)=\left(1+\dfrac{\alpha}{E^{*}}\right)\int_{0}^{x}\dfrac{dy}{1-\dfrac{E(y)}{E^{*}}},\label{Eq.5}
\end{equation}
where $E(y)$ is obtained from the inverse relation of the function $\frac{B(E)}{\int_{0}^{\infty}dE'B(E')}$.\\
In our case, using the distribution function Eq.(\ref{Eq.2}) and the Boltzmann factor Eq.(\ref{Eq.3}), we find the following form for $E(y)$ and $E^{*}$:
\begin{equation}
    E(y)=\dfrac{y^{-x}-1}{x},\qquad E^{*}=-\dfrac{1}{x}.\label{Eq.6}
\end{equation}
An elementary integration of Eqs.(\ref{Eq.5}), we arrive at the following expressions $s(x)$ and $u(x)$:
\begin{equation}
    u(x)=x^{x+1},\qquad s(x)=1-x^{x}. \label{Eq.7}
\end{equation}
Finally, we can write the entropy $S_{+}$ as:

\begin{equation}
S_{+}=\kappa_{B}\sum_{l=1}^{\Omega}\left(1-p_{l}^{p_{l}}\right).\label{Eq.8}
\end{equation}

If we now properly change $p_{l}$ to $-p_{l}$ in Eq.(\ref{Eq.2}) and following the above procedure, another entropy measure can be obtained defined as:
\begin{equation}
    S_{-}=\kappa_{B}\sum_{l=1}^{\Omega}\left(p_{l}^{-p_{l}}-1\right).\label{Eq.9}
\end{equation}
The entropy measures $S_{+}$ and $S_{-}$ can be expressed as functions of the Boltzmann--Gibbs entropy,
$S_{\mathrm{BG}} = -k_{B}\sum_{l=1}^{\Omega} p_{l}\ln p_{l},$ in the particular case of an equiprobable microstate distribution, $p_{l}=1/\Omega$. In this representation, one finds:

\begin{align}
    &\dfrac{S_{+}}{k_{B}}=S_{BG}-\dfrac{S_{BG}^{2}}{2!}e^{-S_{BG}}+\dfrac{S_{BG}^{3}}{3!}e^{-2S_{BG}}-\dots,\nonumber\\
    &\dfrac{S_{-}}{k_{B}}=S_{BG}+\dfrac{S_{BG}^{2}}{2!}e^{-S_{BG}}+\dfrac{S_{BG}^{3}}{3!}e^{-2S_{BG}}+\dots,
\end{align}
where the above expressions follow from the Taylor series expansion of the functions $s_{\pm}(x)=\pm x(1-x^{\frac{1}{x}})$, with $x=\Omega=e^{S_{BG}}$. The role played by the above entropies in the gravitational regime can be seen in \cite{obregon2010superstatistics},  where Newton's law of gravitation is obtained and corrected by these new entropies. The correction terms introduced by this entropy pair to the Newtonian law of gravitation play an important role when the eigendimensions of the object are close to the Planck length $l_{\rm Pl}$. This connection can be realized by setting the Boltzman-Gibbs entropy $S_{BG}$ as: $S_{BG}= \frac{A}{4l_{\rm Pl}^{2}}$. We proceed by extremizing the entropy measures Eq.(\ref{Eq.8}) and Eq.(\ref{Eq.9}), subject to the constraints of energy conservation $\sum_{l=1}^{\Omega}p_{l}^{p_{l}\pm 1}E_{l}$ and probability $\sum_{l=1}^{\Omega}p_{l}=1$. By using the Lagrange multiplier method, we define the following functionals $\Phi_{\pm}$:
\begin{equation}
    \Phi_{\pm}=\dfrac{S_{\pm}}{k_{B}}-\gamma\sum_{l=1}^{\Omega}p_{l}-\beta\sum_{p_{l}}^{\Omega}p_{l}^{p_{l}\pm 1}E_{l},\label{Eq.11}
\end{equation}
here $\gamma$ and $\beta$ are Lagrange multipliers. If we now proceed to vary the functionals $\Phi_{\pm}$ with respect to $p_{l}$ we obtain the following expressions:
\begin{align}
    &1+\ln(p_{l})+\beta E_{l}(1+p_{l}+p_{l}\ln(p_{l}))=p_{l}^{-p_{l}}\quad \textup{To}\quad S_{+},\\ \nonumber
    &1+\ln(p_{l})+\beta E_{l}(1-p_{l}-p_{l}\ln(p_{l}))=p_{l}^{p_{l}}\quad \textup{To}\qquad S_{-},\label{Eq.12}
\end{align}
As it is possible to observe, there is no analytical way to express the distribution of microstates $p_{l}$, as a function of energy $E_{l}$, for both entropy measures $S_{\pm}$. However, we can give a numerical representation of $p_{l}(E_{l})$, considering that $p_{l}$ has the following form:
\begin{equation}
    P_{\pm}(x)=e_{\pm}(-x)=e^{-x}\sum_{j=0}^{\infty}a_{j}^{\pm}x^{j},\qquad x=\beta E_{l}.\label{Eq.13}
\end{equation}
the development coefficients $a_{j}^{\pm}$ can be set numerically. The first four coefficients are shown in Tab.\ref{Table.1}.
\begin{table}[h!]
\caption{First numerically adjusted development coefficients \( a^{\pm}_{j} \).\\
The superscript identifies each entropy measure.}
\centering
\begin{tabular}{ccc}
\toprule
\( j \) & \( a_{+}^{j} \) & \( a_{-}^{j} \) \\
\midrule
4 & 0.00688475 & -0.016203 \\
3 & -0.12706629 & 0.067023 \\
2 & 0.41223182 & 0.071091 \\
1 & 0.03802564 & -0.451767 \\
0 & 1 & 1 \\
\bottomrule
\end{tabular}
\label{Table.1}
\end{table}
Once the approximate form of the microstate probability distribution function $p_{l}$ is known, we determine in the following the new Hamiltonians $\bar{H}_{\pm}$ which are obtained from the different entropy measures $S_{\pm}$, can be elucidated. Some works in the thermo-statistical context where the distribution function Eq.(\ref{Eq.13}) has been used can be found in \cite{Obregon2018,Obregon,LopezPicon2022}.
\subsection{Effective Hamiltonians \texorpdfstring{$H_{\pm}$}{H}}
To derive effective Hamiltonians from the entropy measures $S_{\pm}$, we assume a statistical identification between energy and probability. In the classical limit, the probability of a system being in a state with phase-space coordinates $(p,q)$ is:
\begin{equation}
    P=e^{-\frac{H(p,q)}{m_{pl}}}=e^{-\bar{H}(p,q)}.\label{Eq.14}
\end{equation}
where $\bar{H}=\frac{H}{m_{pl}}$, is the dimensionless Hamiltonian. In the generalized case, the probability distribution becomes:
\begin{equation}
    P_{\pm}=e^{-\bar{H}}\sum_{j=0}^{\infty}\bar{a}_{j}^{\pm}\bar{H}^{j},
\end{equation}
where now the coefficients of development $[\bar{a}_{j}^{\pm}]=1$ are dimensional.  On the other hand, we can consider the statistical effects of both entropy measures, by proposing that the right-hand side of the above expression has the form:
\begin{equation}
    e^{-\bar{H}}\sum_{j=0}^{\infty}\bar{a}_{j}^{\pm}\bar{H}=\exp(-\bar{H}_{\pm}).\label{Eq.16}
\end{equation}
If we now take the natural logarithm of both sides of the expression and by making use of the serial expression of $\ln(1+x)=x-\frac{x^{2}}{2!}+\frac{x^{3}}{3!}+...$, with $\vert x\vert\ll1$ we get for example for $H_{+}$:
\begin{equation}
    \bar{H}_{+}=(1-\bar{a_{1}}^{+})\bar{H}+\frac{1}{2}\bar{H}^{2}(\bar{a_{1}}^{+2}-2\bar{a_{2}}^{+})-\frac{1}{3}\bar{H}^{3}(\bar{a_{1}}^{+3}-3\bar{a_{1}}^{+}\bar{a_{2}}^{+}+3\bar{a_{3}}^{+})\dotsb,\label{Eq.17}
\end{equation}
Similarly for $\bar{H}_{-}$:
\begin{equation}
\bar{H}_{-}=(1-\bar{a_{1}}^{-})\bar{H}+\frac{1}{2}\bar{H}^{2}(\bar{a_{1}}^{-2}-2\bar{a_{2}}^{-2})-\frac{1}{3}\bar{H}^{3}(\bar{a_{1}}^{-3}-3\bar{a_{1}}^{-}\bar{a_{2}}^{-}+3\bar{a_{3}}^{-})\dotsb,\label{Eq.18}
\end{equation}
The fundamental underlying difference between the two Hamiltonians comes from each entropy measure, i.e., they are differentiated by the value of their coefficients $\bar{a}_{j}^{\pm}$, which are shown in Tab.\ref{Table.1}. Another important feature of the $\bar{H}_{\pm}$ Hamiltonians is that they effectively emerge as a function of the low-energy $\bar{H}$ Hamiltonian, which can ultimately be expressed in the form $\bar{H}=\frac{\bar{k}^{2}}{2}+V(\bar{q})$, where except in terms of dimensionality, $\bar{k}$ is the linear momentum and $V(\bar{q})$ is the interaction potential.  In this form, the quantum effects are encoded by the development coefficients. In \cite{Bizet_2023} the free particle case $V=0$ was studied and examined in the context of the Generalized Uncertainty Principle (GUP), where the Hamiltonians $\bar{H}_{\pm}$ can be taken to the form $\bar{H}_{\pm}=\frac{\bar{p}_{\pm}^{2}}{2}$, in this form, new anticonmutation relations can be obtained in the configuration space.

Now, for the purpose of inspecting the resulting effective dynamics of the Hamiltonian $\bar{H}_{\pm}$ in the context of black holes, we choose to use a normalized version of $\bar{H}_{\pm}$, i.e., if we divide both expressions by the quantity $(1-a^{\pm}_{1})$ we have:
\begin{equation}
\mathcal{H}_{\pm}=\dfrac{\bar{H}_{\pm}}{1-\bar{a_{1}}^{\pm}}=\bar{H}+\dfrac{(\bar{a_{1}}^{\pm})^{2}-2a_{2}^{\pm}}{2(1-a_{1}^{\pm})}\bar{H}^{2}+\dotsb, \label{reducedHamiltonian}
\end{equation}
In the following we introduce the parameter $\alpha_{\pm}=\frac{(\bar{a_{1}}^{\pm})^{2}-2a_{2}^{\pm}}{2(1-a_{1}^{\pm})}$, such parameter models the corrections introduced to the previous Hamiltonians. If we consider the values of Tab.(\ref{Table.1}), it is possible to see that $\alpha_{+}<0$ and $\alpha_{-}>0$.  The physical implications of the above parameter can be reviewed in \cite{Bosso2021,Bizet_2023}. In the next section, we use the newly derived Hamiltonians in the high-energy regime in order to study the corrections introduced to the black hole metric whose components are elements of a phase space. The $\bar{H}$ Hamiltonian we will employ is described in Ashtekar-Barbero variables\cite{Ashtekar_2018}, where a phase space can be elucidated to model the inner of a spherically symmetric black hole. With respect to the causal structure of the black hole, this can be considered as a Kantowski-Sachs cosmological metric \cite{Kelly_2020,Gambini_2020,bojowald2020blackholemodelsloopquantum}.
\section{Effective classical dynamics inside the black hole}
\label{sec:dynamics}
In this section, we derive the modified equations of motion for a Schwarzschild black hole, incorporating the effective Hamiltonians 
$\mathcal{H}_{\pm}$ (see Eq.\eqref{reducedHamiltonian}), obtained in the previous section. Since these corrections become relevant near the Planck scale, we retain terms up to second order in the dimensionless Hamiltonian 
$\bar{H}$. 
\subsection{Classical Schwarzschild Interior in Ashtekar-Barbero Variables}
We begin by reviewing the usual classical dynamics for a Schwarszchild-type black hole in the Ashtekar-Barbero variables.  The internal metric for this black hole in terms of the Ashtekar-Barbero variables is given by:
\begin{equation}
    ds^{2}=-N^{2}(T)dT^{2}+\dfrac{p_{b}^{2}(T)}{L_{0}^{2}\vert p_{c}(T)\vert}dr^{2}+\vert p_{c}(T)\vert d^{2}\Omega_{2},\label{eq.22}
\end{equation}
where $N(T)$ is the Lapse function that plays the role of a Gauge parameter in the ADM( Arnowitt-Deser-Misner) formalism. Here $T$ is a generic time parameter and $r$ a spatial variable inside the black hole. The variables $b$, $c$, $p_{b}$ and $p_{c}$, are the Ashtekar-Barbero $A^{i}_{a}$ connection components and the densitized triad $\xi^{a}_{i}$, ie:
\begin{eqnarray}
    A^{i}_{a}\tau_{i}dx^{a}=\frac{c}{L_{0}}\tau_{3}dr+b\tau_{2}d\theta-b\tau_{1}\sin\theta d\phi+\tau_{3}\cos\theta d\phi,\\
    \xi^{a}_{i}\tau_{i}\partial_{a}=p_{c}\tau_{3}\sin\theta\partial_{r}+\frac{p_{b}}{L_{0}}\tau_{2}\sin\theta\partial_{\theta}-\frac{p_{b}}{L_{0}}\tau_{1}\partial_{\phi},
\end{eqnarray}
where $b(T),c(T)$ are variables of a configuration space and $p_{b}(T),p_{c}(T)$ are their associated canonical variables. Here, $L_{0}$, is a fiducial length required for regularizing the symplectic structure. The non-vanishing Poisson brackets are:
\begin{equation}
    \lbrace b,p_{b}\rbrace=G\gamma,\quad \lbrace c,p_{c}\rbrace=2G\gamma, \label{Eq.23}
\end{equation}
where $\gamma$ is the Barbero-Immirzi \cite{Thiemann_2007} parameter, which can be fixed through black hole entropy calculations in Loop Quantum Gravity (LQG). In our analysis, we adopt $\gamma=\frac{\ln(2)}{\pi \sqrt{3}}$.
\vspace{0.1mm}

\begin{figure}[ht]
    \centering
    \includegraphics[width=1\linewidth]{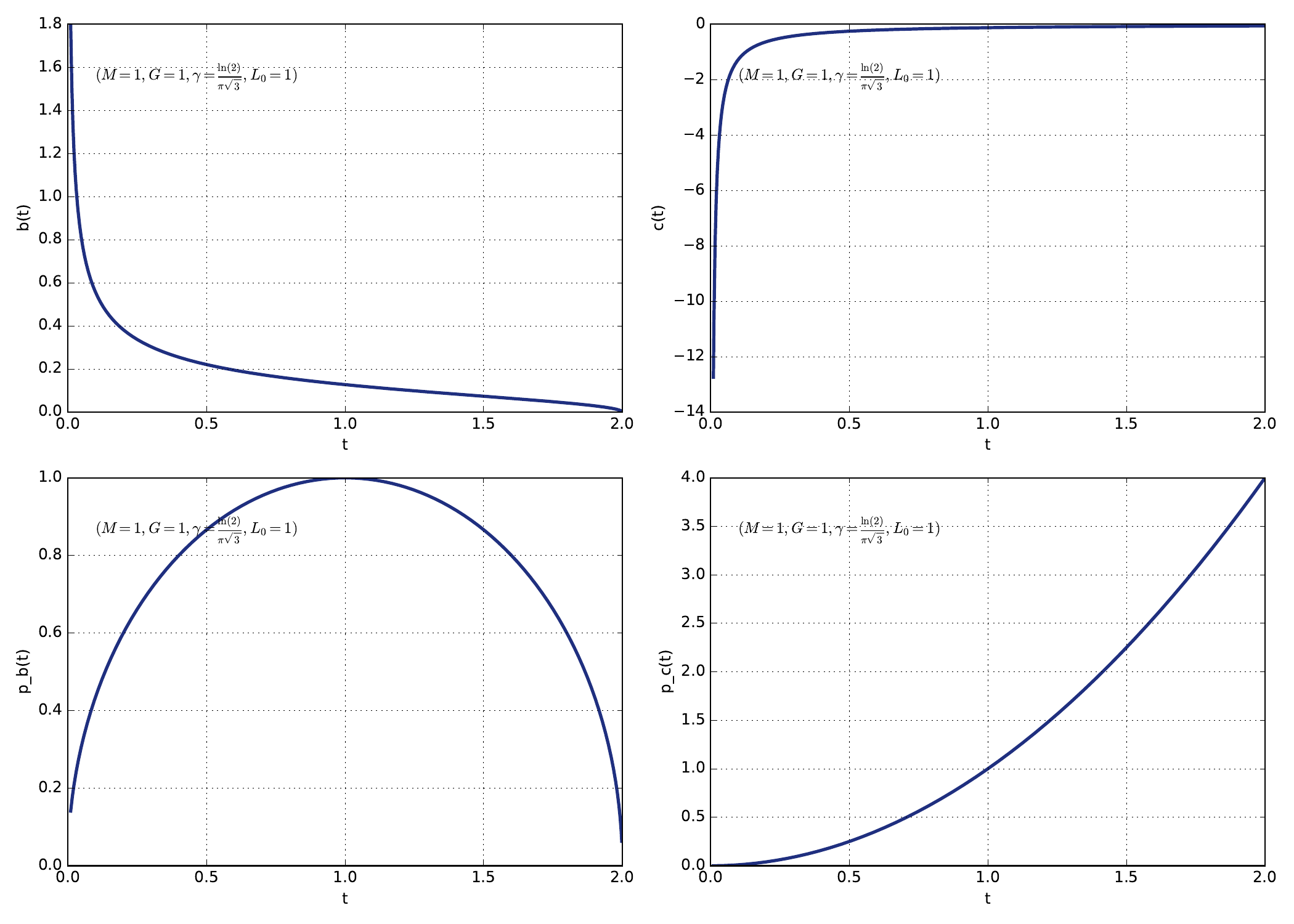}
    \caption{Graphical behavior of the internal solutions to the black hole given by $b,c,p_{b}$ and $p_{c}$ as a function of Schwarszchild time $t$ in the interval $t\in (0,2GM]$.}
    \label{fig:3}
\end{figure}

The classical Hamiltonian (constraint) adapted for this model taking into account its symmetries becomes \cite{PhysRevD.78.044019}:
\begin{equation}
   H=-\frac{8\pi N(T)}{\gamma^{2}}\frac{sgn(p_{c}(T))}{\sqrt{\vert p_{c}(T)\vert}}\left[2bcp_{c}+(b^{2}+\gamma^{2})p_{b}\right].\label{Eq.24}
\end{equation}
The equations of motion resulting from the above Hamiltonian $\dot{f}=\lbrace f,H\rbrace$, for the phase space formed by the canonical variables $b,c,p_{b},p_{c}$, can be decoupled if we make the choice of the following Lapse function $N(T)$ \cite{Melchor2024, Bosso2021}:
\begin{equation}
    N(T)=\frac{\gamma sgn(p_{c}(T))\sqrt{\vert p_{c}(T)\vert}}{16\pi G}.\label{Eq.25}
\end{equation}
The above choice of the Lapse function reduces the Hamiltonian to the following form:
\begin{equation}
    H=-\frac{1}{2G\gamma^{2}}\left[(b^{2}+\gamma^{2})p_{b}+2cbp_{c}\right].\label{Eq.26}
\end{equation}
Other possible choices of the Lapse function $N(T)$ applied in different approximations can be seen in \cite{Corichi2016,Ashtekar2006,modesto2008black}. The respective evolution equations of the phase space formed by $b,c,p_{b}$ and $p_{c}$ are:
\begin{align}
    &\frac{db(T)}{dT}=\lbrace b,H\rbrace=-\frac{b^{2}+\gamma^{2}}{2G\gamma},\quad
    \frac{dc(T)}{dT}=\lbrace c,H\rbrace=-\frac{2bc}{G\gamma},\nonumber \\
    &\frac{dp_{b}(T)}{dT}=\lbrace p_{b},H\rbrace=\frac{bp_{b}+cp_{c}}{G\gamma},\quad\frac{dp_{c}(T)}{dT}=\lbrace p_{c},H\rbrace=\frac{2p_{c} b}{G\gamma}. \label{Eq.27}
\end{align}
Notably, the evolution equation for 
$b(T)$ is decoupled and can be solved independently. The solutions for the remaining variables follow by substitution. The general solutions are:
\begin{align}
    & b(T)=-\gamma\tan\left(\frac{T}{2G\gamma}\right),\quad c(T)= C_{3}\cos^{-4}\left(\frac{T}{2G\gamma}\right),\nonumber\\
    &p_{b}(T)=C_{2}\cos\left(\frac{T}{2G\gamma}\right)\sin\left(\frac{T}{2G\gamma}\right),\quad p_{c}(T)=C_{4}\cos^{4}\left(\frac{T}{2G\gamma}\right). \label{Eq.28}
\end{align}
where $C_{i}$ with $(i=2,3,4)$ are integration constants. To interpret these solutions geometrically, we relate the coordinate $T$ to Schwarzschild time $t$ by identifying 
$p_{c}(t)=t^{2}$, which corresponds to the areal radius of the two-sphere:
\begin{equation}
    \cos^{2}\left(\frac{T}{2G\gamma}\right)=\frac{t}{\sqrt{C_{4}}}.\label{Eq.29}
\end{equation}
If we now use the elementary trigonometric relations $\tan(x)=\pm\frac{\sqrt{1-cos^{2}(x)}}{\cos(x)}$ and $\sin(x)=\pm\sqrt{1-cos^{2}(x)}$, the solution set given by Eqs.(\ref{Eq.27}), in terms of the Schwarszchild time and further choosing the integration constant $C_{4}=4G^{2}M^{2}$, we have:
\begin{align}
    &b(t)=\pm\gamma\sqrt{\frac{2GM}{t}-1},\quad c(t)=\mp\frac{GM\gamma L_{0}}{t},\nonumber\\
    &p_{b}(t)=L_{0}t\sqrt{\frac{2GM}{t}-1},\quad p_{c}(t)=t^{2}.\label{Eq.30}
\end{align}
The rest of the integration constants can be fixed by virtue of the spatial components of the Eq.(\ref{eq.22}) metric. These expressions reproduce the interior Schwarzschild geometry written in Kantowski-Sachs form:
\begin{equation}
    ds^{2}=-N^{2}(t)dt^{2}+X^{2}(t)dr^{2}+Y^{2}(t)d^{2}\Omega_{2}.
    \end{equation}
In terms of phase space variables:
\begin{align}
   & g_{rr}(t)=X^{2}(t)=\frac{p_{b}^{2}(t)}{L_{0}^{2}p_{c}(t)}=\frac{2GM}{t}-1,\quad g_{\Omega\Omega}(t)=Y^{2}(t)=p_{c}(t)=t^{2},\nonumber\\
   &g_{tt}(t)=-N^{2}(t)=\left(\frac{2GM}{t}-1\right)^{-1}. \label{Eq.32}
\end{align}
 The geometric nature of the solution can be obtained through the Kretschmann scalar $K$ defined in the Ashtekar-Barbero variables as \cite{rastgoo2022probing}:
\begin{equation}
    K=\dfrac{12(b^{2}+\gamma^{2})^{2}}{\gamma^{4}p_{c}^{2}(t)}.\label{eq.33}
\end{equation}
As expected, when we approach the singularity $t\rightarrow 0$, the canonical variable $p_{c}\rightarrow 0$ and the Kretschmann scalar diverges $K\rightarrow \infty$, this implies that any curvature inavariant diverges at the singularity having a physical singularity. On the other hand, when approaching the horizon from inside the hole $t\rightarrow 2GM$, the radius of the 2-sphere tends to $p_{c}\rightarrow 4G^{2}M^{2}$ and the Kretschmann scalar remains finite in such a limit $K\rightarrow \frac{3}{4G^{4}M^{4}}$, this means that the horizon is a coordinate singularity that can be removed by a coordinate transformation.  The graphical behavior of the functions $b(t), c(t), p_{b}(t)$ and $p_{c}(t)$ are shown in Fig.(\ref{fig:3}).
\subsection{Entropy-Corrected Dynamics via Effective Hamiltonians}

Let us now consider the reduced Hamiltonian, $\mathcal{H}_{\pm}$, given in Eq.\eqref{reducedHamiltonian}. This can be expanded up to second order, so that the Hamiltonian associated with the generalized entropies in Eqs. (\ref{Eq.8}) and (\ref{Eq.9}) takes the form:
\begin{equation}
    \mathcal{H}_{\pm} = \bar{H} + \alpha_{\pm} \bar{H}^2.
\end{equation}
The effective Hamiltonian corresponding to the reduced one can then be obtained by substituting Eq.\eqref{reducedHamiltonian} into the expression above:

\begin{equation}
    \mathcal{H}_{\pm} = -\frac{1}{2G\gamma^{2}}\left((b^{2}+\gamma^{2})p_{b}+2cbp_{c}\right) 
    + \frac{\alpha_{\pm}}{4G^{2}\gamma^{4}}\left((b^{2}+\gamma^{2})p_{b}+2cbp_{c}\right)^{2}. \label{Eq.35}
\end{equation}

We emphasize at this point that the previously chosen time function $N(T)$ is maintained in the following developments. We can now analyze how the dynamics inside the black hole is modified by the effective corrections to the reduced Hamiltonian, Eq. (\ref{reducedHamiltonian}).
To do this, we determine the new corrected evolution equations corresponding to the phase space $(b,c,p_{b},p_{c})$ where the evolution parameter is $T$. Proceeding in the same way as in the previous section, we have:
\begin{align}
    &\dfrac{db}{dT}=\lbrace b,\mathcal{H}_{\pm}\rbrace=-\dfrac{b^{2}+\gamma^{2}}{2G\gamma^{2}}\left[1-\dfrac{\alpha_{\pm}}{G\gamma^{2}}\left((b^{2}+\gamma^{2})p_{b}+2cbp_{c}\right)\right],\nonumber\\
    &\dfrac{dc}{dT}=\lbrace c,\mathcal{H}_{\pm}\rbrace=-\dfrac{2bc}{G\gamma^{2}}\left[1-\dfrac{\alpha_{\pm}}{G\gamma^{2}}\left((b^{2}+\gamma^{2})p_{b}+2cbp_{c}\right)\right],\nonumber\\
&\dfrac{dp_{b}}{dT}=\lbrace p_{b},\mathcal{H}_{\pm}\rbrace=\dfrac{bp_{b}+cp_{c}}{G\gamma^{2}}\left[1-\dfrac{\alpha_{\pm}}{G\gamma^{2}}\left((b^{2}+\gamma^{2})p_{b}+2cbp_{c}\right)\right],\nonumber\\
&\dfrac{dp_{c}}{dT}=\lbrace p_{c},\mathcal{H}_{\pm}\rbrace=\dfrac{2bp_{c}}{G\gamma^{2}}\left[1-\dfrac{\alpha_{\pm}}{G\gamma^{2}}\left((b^{2}+\gamma^{2})p_{b}+2cbp_{c}\right)\right].\label{Eq.36}
\end{align}
The dynamical system under consideration consists of a set of coupled, highly non-linear differential equations. Nevertheless, the system admits an exact analytical treatment due to a remarkable structural property. All evolution equations share a common multiplicative factor,
\begin{equation}
F = 1-\frac{\alpha_{\pm}}{G\gamma^{2}}\left[(b^{2}+\gamma^{2})p_{b}+2cbp_{c}\right],
\end{equation}
which factorizes universally and cancels out when ratios of time derivatives are taken. As a consequence, relations among the phase-space variables remain directly integrable.

Taking the ratio of the evolution equations for $b$ and $c$ immediately yields
\begin{equation}
\frac{db}{dc}=\frac{b^{2}+\gamma^{2}}{4bc},
\end{equation}
whose solution is
\begin{equation}
c(b)=C_{1}(b^{2}+\gamma^{2})^{2},
\end{equation}
with $C_{1}$ an integration constant. Importantly, this result is independent of the deformation parameter $\alpha_{\pm}$.

The simplification above is ultimately rooted in the conservation of the phase-space combination:
\begin{equation}
H=(b^{2}+\gamma^{2})p_{b}+2cbp_{c},
\end{equation}
which implies that $F$ is constant along the dynamical trajectories and merely rescales the temporal evolution. Consequently, the corrected dynamics differ from the undeformed case only through a shift in the effective frequency, without modifying the functional dependence of the solutions.

Expressing the momenta as functions of $c$, the remaining equations reduce to
\begin{equation}
\frac{dp_{b}}{dc}=-\frac{1}{2}\frac{bp_{b}+cp_{c}}{bc},\qquad
\frac{dp_{c}}{dc}=-\frac{p_{c}}{c},
\end{equation}
whose solutions read
\begin{equation}
p_{c}(c)=\frac{C_{2}}{c},\qquad
p_{b}(c)=\frac{C_{3}}{\sqrt{c}}-\frac{C_{2}\sqrt{C_{1}}}{\sqrt{c}}
\sqrt{\sqrt{\frac{c}{C_{1}}}-\gamma^{2}},
\end{equation}
with $C_{2}$ and $C_{3}$ constants of motion.

Substituting these expressions back into the equation for $\dot c$, one finds
\begin{equation}
\frac{dc}{dT}=-\frac{2b(c)c}{G\gamma^{2}}
\left(1-\frac{\alpha_{\pm}}{G\gamma^{2}}\sqrt{\frac{C_{3}}{C_{1}}}\right),
\end{equation}
which can be integrated straightforwardly. The solution is
\begin{equation}
c(T)=C_{1}\gamma^{4}\sec^{4}\!\left[
\frac{T}{2G\gamma}\left(1-\frac{\alpha_{\pm}}{G\gamma^{2}}\sqrt{\frac{C_{3}}{C_{1}}}\right)
\right].
\end{equation}

The remaining phase-space variables are then given by
\begin{align}
b(T) &=-\gamma\tan\!\left[
\frac{T}{2G\gamma}\left(1-\frac{\alpha_{\pm}}{G\gamma^{2}}\sqrt{\frac{C_{3}}{C_{1}}}\right)
\right],\\
p_{c}(T) &=\frac{C_{2}}{C_{1}\gamma^{4}}
\cos^{4}\!\left[
\frac{T}{2G\gamma}\left(1-\frac{\alpha_{\pm}}{G\gamma^{2}}\sqrt{\frac{C_{3}}{C_{1}}}\right)
\right],\\
p_{b}(T) &=\frac{C_{3}}{\gamma^{2}\sqrt{C_{1}}}
\cos^{2}\!\left[
\frac{T}{2G\gamma}\left(1-\frac{\alpha_{\pm}}{G\gamma^{2}}\sqrt{\frac{C_{3}}{C_{1}}}\right)
\right]
+\frac{C_{2}}{\gamma}
\sin\!\left[
\frac{T}{G\gamma}\left(1-\frac{\alpha_{\pm}}{G\gamma^{2}}\sqrt{\frac{C_{3}}{C_{1}}}\right)
\right].
\end{align}

Note that in the absence of corrections $\alpha_{\pm}\rightarrow 0$, this is equivalent to the low-energy regime, the solutions $b(T),c(T),p_{b}(T),p_{c}(T)$ reduce to the set of equations given in Eq.\eqref{Eq.28}. This allows to fix the integration constants $C_{1}$ and $C_{2}$ in the form:
\begin{equation}
    C_{1}=\dfrac{ L_{0}}{4GM\gamma^{3}}, \quad C_{2}=\gamma GML_{0}.
\end{equation}
Now, to fix the constant of integration $C_{3}$, we note that it plays the role of a phase term in argument of the functions of $(b,c,p_{b},p_{c})$. In this form, we can set $C_{3}$ such that, the argument of the trigonometric functions are dimensionless. The above, fixes $C_{3}$ in the form:
\begin{equation}
    C_{3}=\dfrac{\gamma G L_{0}}{4M}.
\end{equation}
Note that, the units of $C_{3}$ are $[C_{3}]=[L]^{2}$. In the following, we discuss the internal dynamics classical effective of the black hole when the quantum corrections modulated by the $\alpha_{\pm}$ parameter and encoded in the solutions $(b , c,p_{b},p_{c})(t)$ are considered in terms of the Schwarszchild time $t$.
\subsection{Interpretation and Geometric Implications}
\begin{figure}[ht]
    \centering
    \includegraphics[width=1\linewidth]{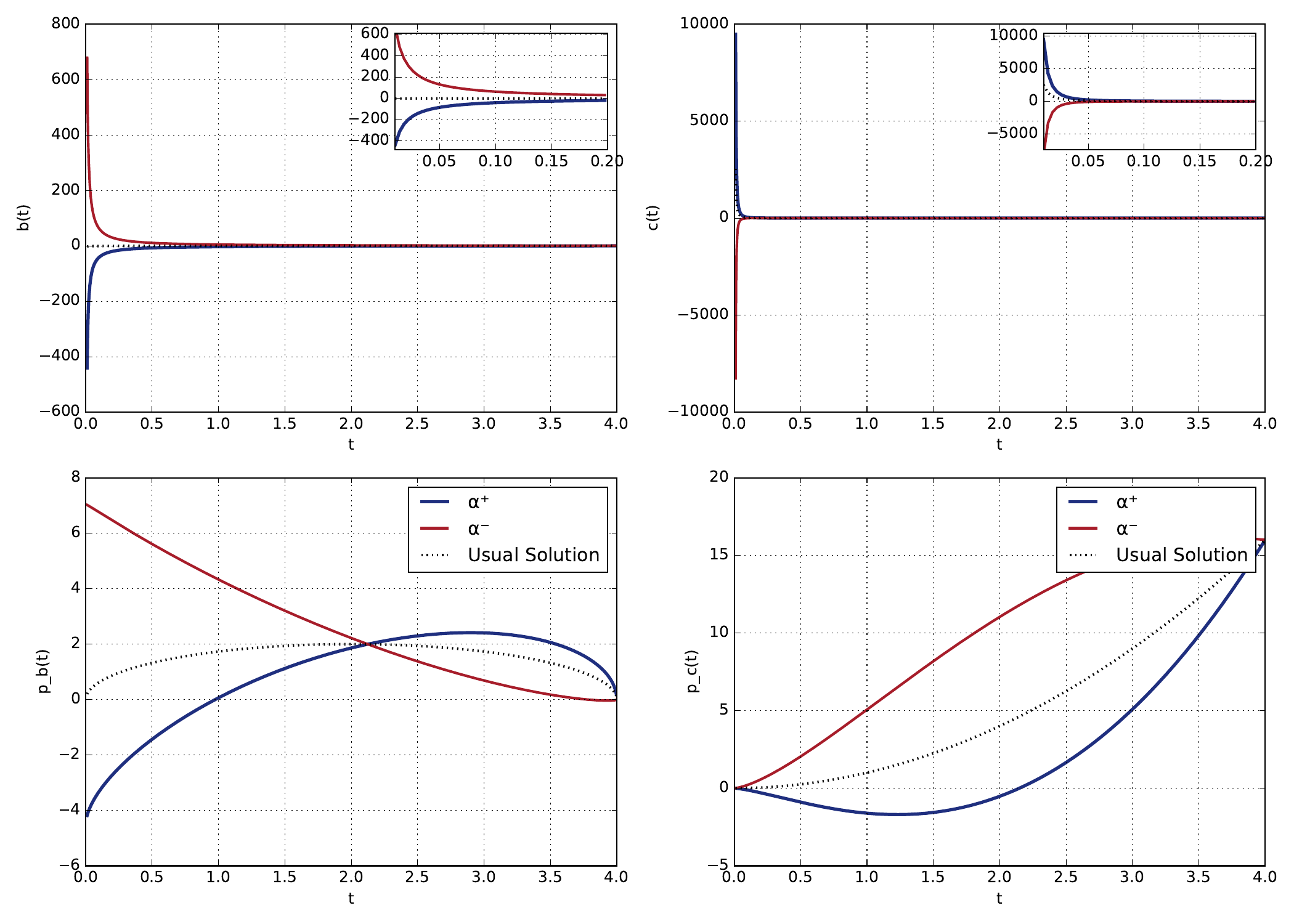}
    \caption{Graphical representation of the phase-space variables $(p_{b},p_{c},b,c)$ as functions of Schwarzschild time $t$. Quantum corrections associated with the entropies $S_{\pm}$	
  are encoded in the parameters $\alpha_{+}=-0.36$ and $\alpha_{-}=0.56$, as given in Table~\ref{Table.1}. The mass of the black hole is set to $M=2$. Dashed lines indicate the classical (uncorrected) solutions, while solid lines correspond to the corrected dynamics.}
    \label{fig:2}
\end{figure}

To connect with the internal geometrical structure of the black hole, we choose the relation between the generic time $T$ and the Schwarszchild time $t$ in the form:
\begin{equation}
    \cos^{2}\left(\dfrac{T}{2G\gamma}\right)=\dfrac{t}{2GM}.
\end{equation}
While the above fixes the size of the 2-sphere in the form $p_{c}=t^{2}$ in the absence of quantum corrections to the internal dynamics of the black hole, in our case, we can consider that $\alpha_{\pm}$ is a small parameter satisfying $\vert\alpha_{\pm}\vert\ll 1$.  In this form, the above solutions can be expanded to first order in $\alpha_{\pm}$ and in terms of the Schwarszchild time we have:
\begin{align}
    &b(t)=-\gamma\sqrt{\dfrac{2GM}{t}-1}+\alpha_{\pm}\dfrac{4GM\arccos\left(\sqrt{\dfrac{t}{2GM}}\right)}{t},\nonumber\\
    &c(t)=\dfrac{GM\gamma L_{0}}{t^{2}}-\alpha_{\pm}\dfrac{\dfrac{L_{0}\gamma}{GM}\arccos\left(\sqrt{\dfrac{t}{2GM}}\right)\sqrt{\dfrac{2GM}{t}-1}}{t^{2}},\nonumber\\
    &p_{c}(t)=t^{2}+4\alpha_{\pm} t^{2}\sqrt{\dfrac{2GM}{t}-1}\arccos\left(\sqrt{\dfrac{t}{2GM}}\right),\nonumber\\
  &p_b(t)=L_{0}\sqrt{2GM t-t^{2}}+\left(\dfrac{GL_{0}}{16M^{3}\gamma}\right)^{1/2}t\nonumber\\
  &+\alpha_{\pm}\arccos\left(\sqrt{\dfrac{t}{2GM}}\right)\Bigg[\left(\dfrac{GL_{0}}{16M^{3}\gamma}\right)^{1/2}\sqrt{2GM t-t^{2}}-4L_{0}(t-GM)\Bigg].\label{Eq.51}
\end{align}
As expected, the first term of the above expansion corresponds to the classical dynamics inside the hole in the absence of quantum corrections in the Hamiltonian. The graphical behavior of the first-order expanded solutions in $\alpha_{\pm}$ are shown in Fig.(\ref{fig:2}) in contrast to the usual solutions in the absence of quantum corrections. 
\vspace{0.1mm}

The effective Hamiltonian dynamics introduces a boundary to the usual solutions; an upper bound by the entropy measure $S_{-} (\alpha_{-})$ and a lower bound introduced by the corrections due to $S_{+}(\alpha_{+})$. Although this behavior holds for some of the variables, for $p_{b}$ it is more subtle. Note that the value of the canonical variable is finite as we approach the singularity from the event horizon of the black hole ($t\rightarrow 0$). In this case, $p_{b}$ takes the following value:
\begin{equation}
    p_{b}(t\rightarrow 0)=4GM L_{0}\alpha_{\pm}.
\end{equation}

The above represents the area $A_{r\theta}, A_{r,\phi}$ of the section of the 2-sphere $\mathbb{S}^{2}$, subtended by the radial component in the polar $\theta$ direction and in the azimuthal $\phi$ direction, given by:
\begin{equation}
    A_{r\theta}=A_{r\phi}=2\pi\vert p_{b}\vert= 8\pi GML_{0}\vert\alpha_{\pm}\vert.
\end{equation}
This means that the black hole singularity is effectively bounded by two circumferences: one in the latitudinal direction with area $A_{r\theta}$ , and another in the longitudinal direction with area $A_{r\phi}$. These areas are not fixed but vary explicitly with the black hole mass $M$, and the quantum correction parameter $\alpha_{\pm}$. This reflects a quantum-induced regularization of the classical singularity, leading to a finite, anisotropic core structure instead of a point-like divergence. Note that the usual case recovers when $\alpha_{\pm}\rightarrow 0$. On the other hand, when we approach the event horizon from inside the black hole $(t\rightarrow 2GM)$ the variable $p_{b}$ tends to the value:
\begin{equation}
    p_{b}(t\rightarrow2GM)=G\left(\dfrac{GL_{0}}{4M\gamma}\right)^{1/2}.
\end{equation}
This implies that the cross-sectional $A_{r\theta}$ and $A_{r\phi}$, remain finite at the black hole’s event horizon, indicating that the horizon corresponds to a geometrically smooth region in the effective description. Notably, this smoothness exhibits an inverse dependence on the black hole mass $M$ and disappears when the mass of the hole increases. Such regular behavior emerges from the linear correction in $p_{b}$ present in the effective Hamiltonian (see Eqs.\eqref{Eq.51}), a feature entirely absent in the standard classical theory. 

If we now consider the surface area of the 2-sphere denoted by $A_{\theta\phi}$ and defined as:
\begin{equation}
    A_{\theta\phi}=\pi\vert p_{c}\vert.
\end{equation}
The behavior of the surface area of the 2-sphere at the singularity $(t\rightarrow 0)$ and at the event horizon $(t\rightarrow 2GM)$ are of the form:
\begin{equation}
    A_{\theta\phi}=0,\quad \textup{when}\quad t\rightarrow 0,\quad A_{\theta\phi}=4G^{2}M^{2}\pi,\quad \textup{when}\quad t\rightarrow2GM.
\end{equation}
However, this is not the whole story. Note that in particular, the correction introduced by the entropy measure $S_{+}$ in our effective model is negative $(\alpha_{+}<0)$ (see Fig.(\ref{fig:2}), for $p_{c}(t)$, blue curve). There is a point on the $t^{*}$ axis where $p_{c}(t^{*})=0$, this corresponds to the solution of the transcendental equation:
\begin{equation}
    (t^{*})^{2}+4\alpha_{+}(t^{*})^{2}\left[\sqrt{\dfrac{2GM}{t^{*}}-1}\right]\arccos\left(\sqrt{\dfrac{t^{*}}{2GM}}\right)=0.
\end{equation}
A first root of the above transcendental equation occurs when $t^{*}=0$. Another solution can be found using a numerical method by varying the constant $GM$. These roots fix the length scale at which quantum effects take place, since these roots depend on the correction parameter $\alpha_{+}$. In this regime, the area of the two-spheres $A_{\theta\phi}$, collapses to zero, signaling the vanishing of the angular directions, while the transverse components $A_{r\theta}=A_{r\phi}$, remain finite. This anisotropic behavior provides clear evidence for the formation of a narrow throat or cigar-like structure inside the black hole, characterized by a residual radial extension of size  $p_{b}(t^{*})$. Similar features appear in the effective dynamics of loop quantum cosmology, particularly in Bianchi models, where the geometry near the bounce becomes highly anisotropic\cite{Bojowald_2005}.

Notably, when the apparent singularity located $t=t^{*}$ is crossed, the geometry transitions into a new spacetime region where the metric components $g_{tt}$ and $g_{rr}$ reverse their signs—i.e., $g_{tt}>0$ and $g_{rr}<0$. This signature change has been previously observed in effective models of black hole interiors in loop quantum gravity\cite{Ashtekar2006,modesto2006loop}, and it signals a smooth extension through what was classically a singularity.

\begin{figure}[ht]
    \centering
    \includegraphics[width=1\linewidth]{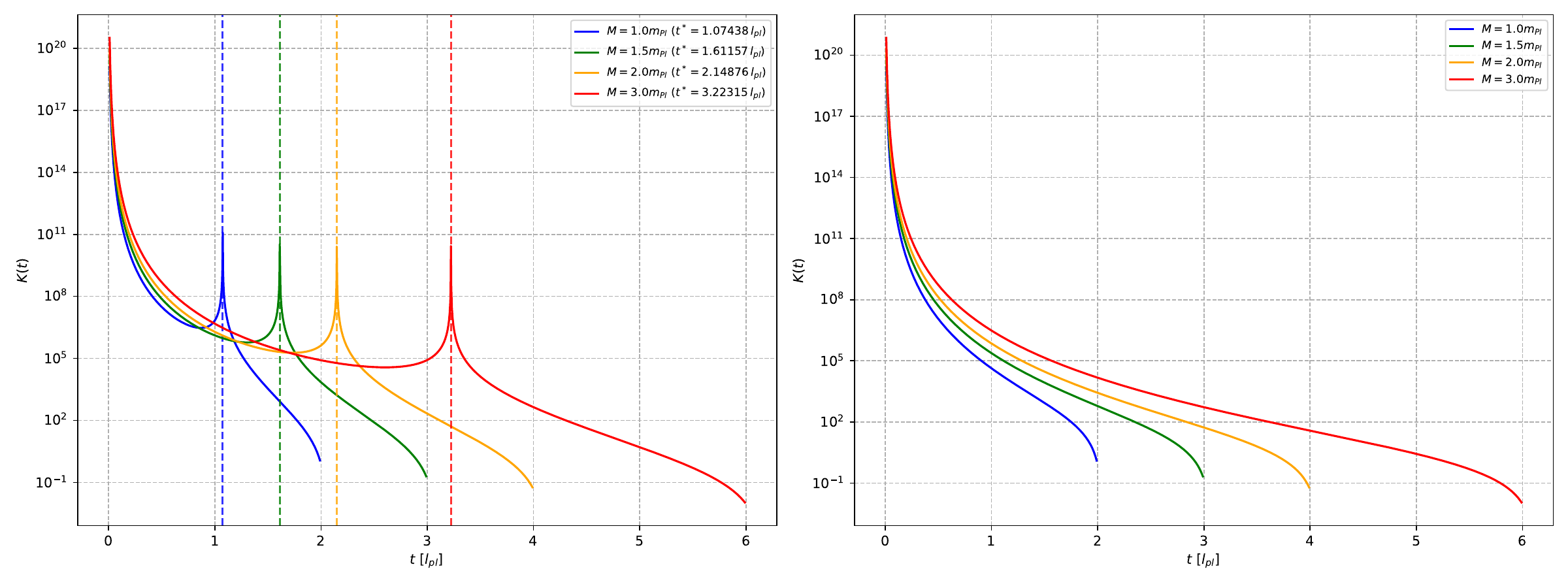}
    \caption{The left panel shows the curvature $K(t)$ for the entropy measure $S_{+}$	
 , with varying hole mass in Planck units $m_{pl}$	
  and the corresponding divergence points $t^{*}$ in $l_{pl}$. The right panel shows $K(t)$ for $S_{-}$	
  under the same mass variations, where the curvature remains finite for $t\in (0,2GM]$.}
    \label{fig: K(t)}
\end{figure}

A key diagnostic of the interior geometry is provided by the Kretschmann scalar $K$ (see Eq.\eqref{eq.33}). In our effective description, there exists a point $t=t^{*}$, where $p_{c}(t^{*})=0$, which arises solely from the quantum corrections associated with $S_{+}$, for $\alpha_{+}<0$. In contrast, for $S_{-}$,  with $\alpha_{-}>0$, this behavior is absent, and the Kretschmann scalar remains finite throughout the interval $t\in(0,2GM]$. The divergence at $t=t^{*}$ for $S_{+}$, indicates that the transition to the interior of the anisotropic core is not entirely smooth, signaling a localized region of extreme curvature. This feature reflects the significant impact of the entropic corrections encoded in $S_{+}$, on the effective geometry, suggesting that the inner structure of the core is more sensitive to quantum effects than in the 
$S_{-}$ case. Such divergences may have important consequences for the physical interpretation of the core dynamics and the propagation of fields in the innermost regions. In Fig.\ref{fig: K(t)}, the behavior of $K(t)$ for the different entropy measures $S_{+}$ and $S_{-}$ is shown, illustrating the contrasting effects of the two entropic corrections on the curvature profile.
\vspace{1mm}

Based on these results, it is important to note that the effects of curvature and its transitions are noticeable at Planck scales. In comparison with a black hole whose mass is of the order of one solar mass $M_{\odot}\sim 10^{30}kg\sim 10^{38}m_{pl}$. The Schwarzschild radius $R_{s,\odot}=\dfrac{2GM}{c^{2}},$ considering a black hole with one solar mass has a size $R_{s,\odot}\sim 10^{38}l_{pl}$. That is, when using the entropy measure $S_{+}$, the characteristic size of the inner core is of the order $t^{*}\sim10^{38}l_{pl}\sim R_{s,\odot}$. This value corresponds to the Schwarzschild radius of a solar-mass black hole and remains indistinguishable for black holes whose masses lie within the classical (macroscopic) regime. The same conclusion holds when considering the alternative entropy measure $S_{-}$.

\section{Conclusions}
In this work, we have investigated the effective dynamics of the Schwarzschild black hole interior using entropy-deformed Hamiltonians derived from generalized entropic measures within the superstatistical framework. By solving the modified equations of motion analytically, we have shown that the inclusion of entropic corrections leads to a semiclassical regularization of the classical singularity. The canonical variables remain bounded throughout the entire evolution, in sharp contrast with the unphysical divergences of the standard Schwarzschild interior.

The geometric interpretation of the corrected dynamics reveals an anisotropic core characterized by cigar-like features. Specifically, the transverse area of the internal 2-sphere, $A_{\theta\phi}$, collapses to zero at the core, whereas the longitudinal areas $A_{r\theta}$ remain finite and proportional to $\vert\alpha_{\pm}\vert$. This residual extension in the radial directions ensures that the singularity is replaced by a finite throat rather than a pointlike divergence, in agreement with loop-inspired models of black hole interiors. At the event horizon, both longitudinal and transverse areas remain finite. A central outcome is the behavior of the Kretschmann scalar: while classically $K\rightarrow\infty$ as $p_{c}\rightarrow 0$, for the entropy measure $S_{-}$, ($\alpha_{-}>0$), this finiteness persists throughout the interval $t\in (0,2GM]$, indicating a completely regular interior. In contrast, for $S_{+}$ ($\alpha_{+}<0$), a point $t^{*}$ appears where $p_{c}(t^{*})=0$, and $K(t)$ formally diverges, signaling that the transition to the anisotropic core is not entirely smooth. This highlights that the corrections induced by $S_{+}$ introduce localized regions of extreme curvature, while the interior remains free from classical singularities. The contrasting behaviors of $K(t)$ for 
$S_{+}$ and $S_{-}$ are illustrated in Fig.\ref{fig: K(t)}.

Another remarkable feature is the emergence of a smooth signature change across a critical radius, where the temporal and radial components of the metric exchange their causal roles. This transition, which parallels results obtained in loop quantum gravity, indicates the possibility of extending the black hole geometry beyond the classical singularity into a new spacetime region.

Altogether, these findings establish that generalized entropies and their associated Hamiltonian deformations reproduce key features of semiclassical black hole models—bounded variables, regular curvature, anisotropic cores, and signature change—without invoking polymer quantization or discrete structures. Instead, the resolution arises from a statistical-mechanical principle rooted in information theory. Future extensions of this framework could address rotating and charged black holes, explore possible connections with holographic entropy bounds, and investigate observational consequences of the entropy-induced regularization mechanisms.

\label{Conclusions}
\section*{Acknowledgments}
O. O thanks to the grant by the University of Guanajuato CIIC 156/2024 “Generalized Uncertainty Principle, Non-extensive Entropies, and General Relativity”, as well
as the SECIHTI grant CBF2023-2024-2923 “Implications of the Generalized Uncertainty Principle (GUP) in Quantum Cosmology, Gravitation, and its Connection with
Non-extensive Entropies”. J.R.P and O.G would like to express their gratitude to SECIHTI for the support received through the program: "Apoyo de becas nacionales para estudio de posgrado".

We would like to thank Dr Wilfredo Y. Carpio for his invaluable insights and careful discussions, which greatly contributed to clarifying several conceptual aspects of this work.
\newpage
\bibliographystyle{iopart-num}
\bibliography{Bibliography}
\end{document}